# Photon Correlations in Colloidal Quantum Dot Molecules Controlled by the Neck Barrier

*Somnath Koley[1], Jiabin Cui[1†], Yossef. E. Panfil[1], Yonatan Ossia[1], Adar Levi[1], Einav Scharf[1], Lior Verbitsky[1] and Uri Banin[1]\**

[1]Institute of Chemistry and the Center for Nanoscience and Nanotechnology, The Hebrew University of Jerusalem, Jerusalem 91904, Israel.

## Abstract

Fused homodimer Colloidal Quantum Dot Molecules (CQDMs), analogous to homonuclear diatomic molecules, exhibit hybridization of the confined electronic states controlled by the neck girth at the fusion interface. Their constituent "artificial atom" CQDs manifest tunable optoelectronic properties relevant for numerous applications, including in future quantum technologies. Thus, the concept of "nanocrystal chemistry" underlying the CQDMs, greatly enriches the selection of such artificial constructs. Numerous new bright multiexcitonic configurations may form in the CQDM, unlike the multiexcitons in CQDs which are dimmed by strong non-radiative Auger recombination processes. The handle of the neck girth is found to dictate the many-body interplay in CQDMs and determines the photon purity, where in narrow neck girth (weak coupling), the CQDM acts as a multiphoton emitter, while neck filled CQDMs (strong coupling) regain single photon emission characteristics typical of the monomers. The unique attributes of the CQDMs manifesting excellent absorbing and bright tunable excitonic and multiexcitonic states highlights their relevance for potential light emitting applications. In this manuscript, we investigate the charge re-distribution upon optical excitation of various necked homodimer CQDMs using single particle emission spectroscopy. By tuning the hybridization of the electron wavefunction at a fixed center-to-center distance through controlling the neck girth, we reveal two coupling limits. On one hand a "connected-but-confined" situation where neighbouring CQDs are weakly fused to each other manifesting a weak coupling regime, and on the other hand, a "connected-and-delocalized" situation, where the neck is filled beyond the facet size leading to a rod-like architecture manifesting strong-coupling. Either coupling regimes entrust distinct optical signatures clearly resolved at room temperature in terms of photoluminescence quantum yield, intensity time traces, lifetimes, and spectra of the neutral-exciton, charged-exciton, and biexciton states. The interplay between the radiative and non-radiative Auger decays of these states, turns emitted photons from the CQDMs in the "weak-coupling" regime highly bunched unlike CQD monomers, while the antibunching is regained at the "strong-coupling' regime. This behavior correlates with the hybridization energy being smaller than the thermal energy (kT ~25meV) at the "weak-coupling" limit ($\Delta E$~5-10meV), leading to exciton localization suppressing Auger decay. In the neck-filled architectures, the larger hybridization energy ($\Delta E$~20-30meV) leads to exciton delocalization while activating the fast charged and multi-exciton Auger decay processes. This work sets an analogy for the artificial molecule CQDMs with regular molecules, where the two distinct regimes of weak- and strong-coupling correspond to ionic- or covalent- type bonding, respectively.

**INTRODUCTION**

Colloidal Quantum Dots (CQDs) are often considered as artificial atoms and have been widely explored in numerous applications due to their excellent size dependent opto-electronic characteristics alongside their flexible processing from solution.[1-11] In particular, core/shell CQDs have reached near-unity photoluminescence quantum yield (PL QY) and utilizing their high color purity has led to their integration in commercial displays with unprecedented color quality.[12-17] Being able to meet the stringent requirements of this demanding application, sets the stage for utilizing light emission of CQDs in new areas including in lasing, in quantum technologies and in super-resolution imaging to name a few.[18-21] To enrich the selection of CQD based functional materials, "Nanocrystal chemistry" was recently introduced, fusing two core/shell CQD monomers to generate the simplest homodimer CQDM.[22-29] The fusion forms an epitaxial connection between the two neighboring CQDs allowing the electronic wavefunctions to hybridize.[22-29] CQD fusion is also an important parameter in electrical transport through CQD solids.[30-34] In the CQDMs, the neck at the interface between the two composing CQDs plays an important role in the hybridization, tuning the electron tunneling and leading to increased red-shift in the band-edge transition upon neck filling.[23,26] Herein, we study the consequences of this tunable coupling on the emission characteristics of single CQDMs focusing on neutral-, charged- and multi-excitonic states. In CQDs, the fate of charged- and mutiexcitonic states has been the subject of intense scrutiny, as they govern the emission characteristics. For example, on-off blinking of the CQDs emission involves the presence of dimmed charged exciton states and extensive efforts have been devoted to suppress this effect. In CQDMs, there are rich possibilities for localized versus segregated charged- and multi-exciton states involving both emission centers highlighting its potential as a model system to study many-body interactions. The neck engineering provides a flexible handle to control their coupling regime.[26] In turn, this is seen to control the interplay between the radiative and non-radiative Auger decay of the charged and multi-excitonic states.[35-38] Weakly fused dimers manifest localization that decreases the Auger decay rate leading to emissive charged- and bi-exciton states, while under strong fusion the single photon emission characteristics of the CQD monomers is partially regained, due to wavefunction delocalization that triggers the non-radiative Auger decay processes.

The consequences of coupling in Quantum Dot dimers has been explored in MBE grown quantum systems, with a vision to apply the hybridization effects to quantum computing applications.[39-45] Distinct coupling regimes were identified, with analogy to ionic- and covalent-type bonding in naturally occurring molecules. The ionic limit refers to the situation of weak coupling versus the covalent limit of strong coupling. Numerous insights were obtained for MBE grown quantum dot molecules, yet this requires expensive fabrication and these dimers are embedded with the host matrix while requiring cryogenic operation conditions applicable to specific scenarios.[39-45] The CQDMs studied herein, however, exhibit a coupling resolvable at room temperature and ambient conditions, and can be flexibly manipulated as demanded by numerous application scenarios, just like the monomer CQDs. The flexibility of the CQDM synthesis is utilized here, based on the recent progress enabling the control of the neck girth, to explore the optical characteristics of CQDMs at various coupling regimes via single particle spectroscopy.

Single particle spectroscopy revolutionized CQDs research and revealed early-on that the emission from excited CQDs blinks similar to a random telegraph signal, oscillating between the "bright-ON" state to a "dimmed-off" state[46-50]. The on state arises from neutral exciton emission, i.e. radiative electron-hole recombination, while the "dimmed-OFF" state arises from fast non-radiative auger recombination in charged excitons.[35-36, 48-51] Bi-excitons in strongly confined type-I CQDs are also dim due to an efficient non-radiative Auger process, and are typically red-shifted by the attractive charge correlations, however repulsive bi-exciton with blue-shifted emission has also been reported.[52-53] Especially in small CQDs, the "bright-ON" neutral exciton state remains distinct from the other multi-carrier states, maintaining a high photon purity as demanded for single photon emitters (Figure S1).[54] For larger CQDs, the multi-carrier states start to emit and impart mixed-photon-statistics while the photon purity decreases.[55] This is related to the interplay of the radiative recombination and the Auger non-radiative recombination, which is dependent largely on the extent of delocalization and overlap of the charge carrier wavefunctions, and thus governing the emitted photon statistics of colloidal quantum dot systems.[56-63]

CQDMs studied herein, offer a new regime for charged and multiexcitonic configurations where localized versus segregated (delocalized) states are possible and can be manipulated by the neck barrier characteristics. By careful synthetic control of the fusion reaction forming CQDMs constructed from the same CQD core/shell monomers, we maintain the same center-to-center distance while modifying the neck girth from weakly fused to rod-like architectures, presenting two limits for coupling. As will be discussed below, this allows to unravel a transition from weak coupling where the CQDM manifests emission of two localized centers, to the strong coupling limit where delocalization leads to regaining partially single photon emission characteristics akin of monomers.

## RESULTS AND DISCUSSIONS

### CQDM synthesis

The CQDMs studied in this work comprise of the model system of fused dimers of CdSe/CdS core/shell CQDs as described in our earlier studies.[22-27] The choice of CdSe/CdS core/shell CQDs is natural as they have been extensively studied reaching a high level of synthetic control and deep understanding of their optoelectronic properties, and in particular, their charge- and multi-exciton characteristics, serving as a powerful background for this work. CQDs with a small 1.4nm radius core, and moderately thin shell (2.1nm thickness) were chosen, as their dimers manifest hybridization and coupling effects (experimental section, Figure S1). The CdSe/CdS core/shells manifest a quasi type-II band alignment, where the electron with the light effective mass ($m_e^* = 0.1 m_0$) and small core-shell conduction band offset ($\Delta_{CB} \sim 0.1 - 0.3\ eV$) delocalizes to the shell while the hole remains confined in the core ($m_h^* = 0.45 m_0$, $\Delta_{VB} \sim 0.5\ eV$).[64] Synthesis of the CQDMs with varying neck girth was performed by the procedures reported in our prior work, utilizing a template-based method.[22] Briefly, at first, monomer CQDs are bound to the surface of ~200nm diameter silica spheres serving as the template. Following necessary masking by a thin silica shell, a tetrathiol linker molecule is added followed by adding the second CQD batch, thus forming organically linked dimers on the silica sphere surface. The non-fused dimers are then released to the solution via selective etching of the silica by HF and cleaned. Finally, a series of thermo-chemical fusion reactions are performed to achieve fused CQDMs with controlled neck girth. The control of the neck is achieved via changing the fusion reaction temperature and the amount of ligands. Via an unique intra-particle ripening mechanism, the neck diameters are tuned while the CQDs center-to-center distance, core, and surface properties remain unaltered (SI for details).[22-26] Figure 1A,B shows representative High-angle annular dark-field scanning transmission electron microscopy (HAADF-STEM) images and energy dispersive spectroscopy (EDS) mapping for the weakly fused and rod-like CQDMs respectively.

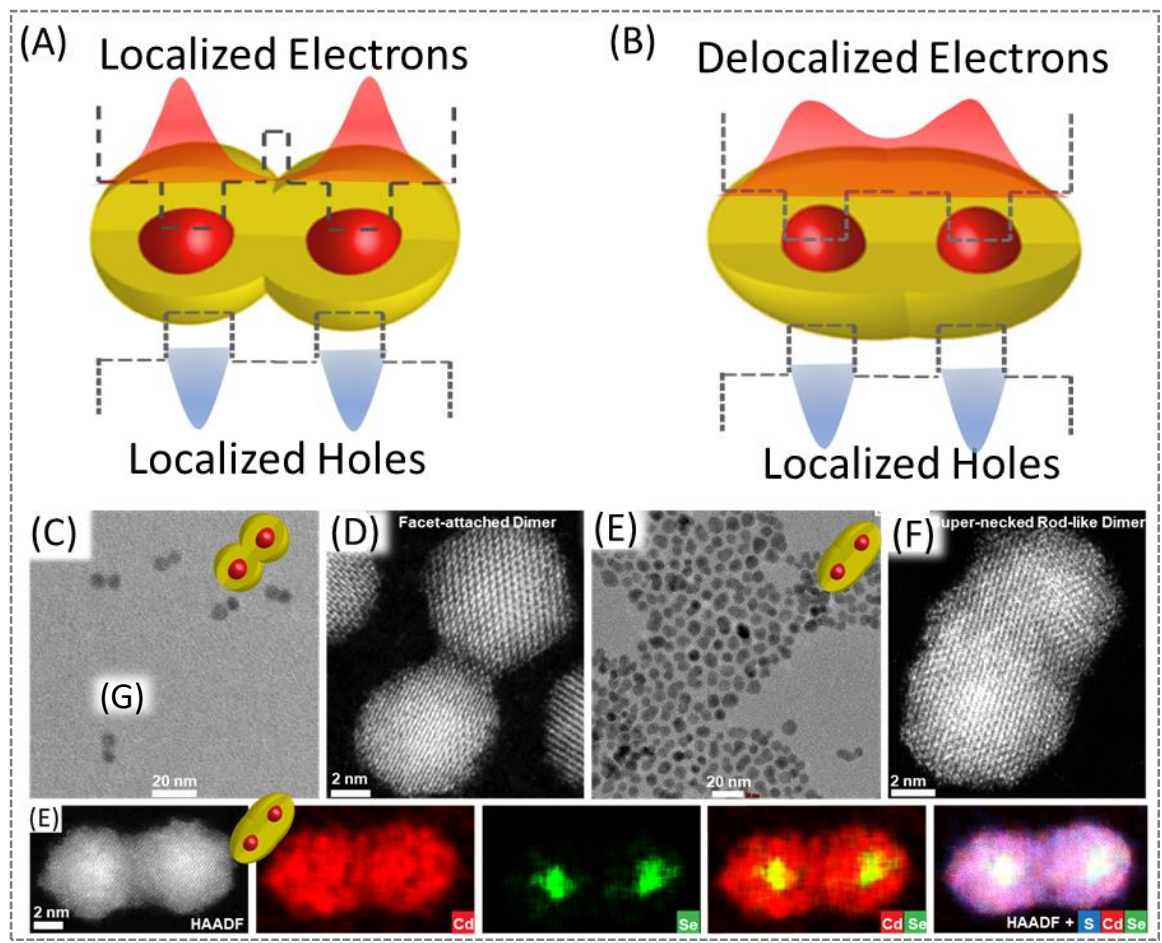

**Figure 1. The model CQDMs with controlled neck girth.** (A,B) Schematics of the potential energy landscape and the electron and hole wavefunctions for the weak-neck and rod-like dimers, respectively. (C,D) TEM, HAADF STEM images of the weakly coupled CQDMs (E,F) same for strongly-coupled rod-like CQDMs, achieved via neck-barrier engineering. (G) HAADF-STEM image and EDS elemental mapping of a strongly coupled CQDM. The cores versus shell regions are clearly identified.

**Optical signatures of the CQDMs in different coupling regimes**

Single particle spectroscopy is ideally suited for this study, as extracting the neutral-, charged-and multiexciton behavior can be garnered individually for different particles. We measured the time-tagged-time-resolved photon stream on single monomer CQDs and CQDMs at the different coupling regimes, allowing to extract the blinking statistics, fluorescence lifetimes, and second order photon correlation ($g^{(2)}$), accompanied by simultaneous spectral information. This was done under low photo-excitation intensity yielding $<N> = \sigma \times J = 0.1$ for monomers, and 0.2 for dimers; where $<N>$ is the average number of excitons, $\sigma$ is the absorption cross-section, and $J$ is the excitation flux density per unit area (Figure 1E, Experimental Procedures).

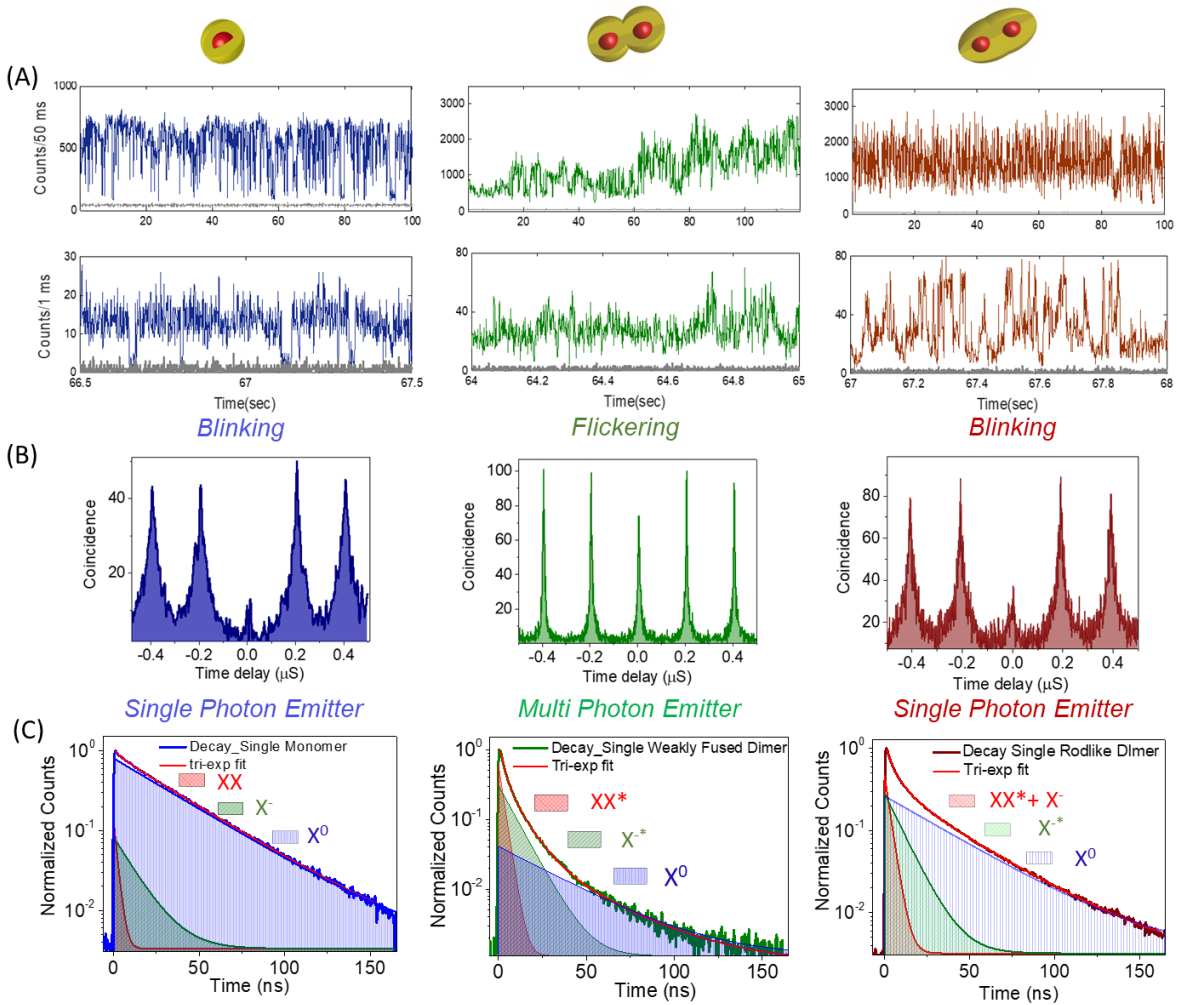

**Figure 2. Comparative single particle emission characteristics of the CQD monomer (left, blue color), weakly coupled CQDM (middle, green color) and strongly coupled CQDM (right, brown color).** (A) Time traces of fluorescence intensities of a single monomer (blue), weakly fused dimer (green), and rod-like dimer (brown) at 50ms (upper panel) and 1ms (lower panel) bin resolution respectively. A well resolved blinking feature common in CQD monomers,

turns into fast flickering for dimers at the weak coupling regime, and is regained to some extent with neck filling at the coupling limit of a rod-like dimer. (B) Second order photon correlation function i.e. $g^{(2)}$ exhibits a strong photon antibunching for the monomer (blue) at τ=0. A strong photon bunching is manifested by the weakly fused dimer (green), which is transformed to an antibunching feature for the neck filled CQD molecules (brown). (C) The time resolved decay profiles (from left to right) for the single monomer, weakly fused dimer, and rodlike dimer, with deconvolution of the lifetime curves for each. Populations of the three distinct states are shown by the different colors; blue- neutral exciton, green- charged exciton, and red-biexciton. (See SI for details).

Figure 2 presents comparative single particle data including fluorescence intensity traces, second order photon correlation traces ($g^{(2)}$), and fluorescence lifetime curves for a representative single monomer CQD alongside single weakly fused and rod-like CQDMs. Distinct differences are seen among these three systems. Starting from the well-studied monomer CQD, its fluorescence intensity traces manifest clear on-off telegraph-like switching behaviour. This fluorescence blinking arises from oscillation between the highly emissive ON state assigned to the neutral exciton ($X^0$) radiative recombination, and the dimmed OFF states, attributed to the non-emissive charged-exciton ($X^-$) and bi-exciton $XX$ states. For the small CdSe/CdS CQDs studied herein, the $X^-$ and $XX$ states are essentially non-emissive due to the efficient competing non-radiative Auger recombination, leading to fast decay reducing their PLQY significantly compared to the near unity PLQY of the neutral exciton state. We note that in these small CQDs, the grey state emission typically assigned to radiative contribution of the $X^-$ state is rare (Figure S2, S3). The $g^{(2)}$ trace for the CQD monomer displays strong antibunching with a contrast between the values at $t = 0 ns$ versus $t = 200 ns$ of less than 0.1, showcasing its good performance as a single photon emitter at room temperature. Moreover, the $g^{(2)}$ contrast directly provides the ratio between the biexciton and exciton PLQY (see below), and considering unity QY of the $X^0$ state, we infer that the biexciton PLQY is less than 10% in accordance with the literature values.[54]

Moving to the weakly fused dimer, its fluorescence intensity time traces differ substantially and instead of the conventional blinking, a flickering behaviour is seen without clear on- or off- states. Note that the emission intensity is significantly higher than that of the monomers. This is due to the nominally doubled absorption cross-section in the CQDM compared with the monomers, and the presence of the two emission centres (Figure 1B, Figure S4). Furthermore, the $g^{(2)}$ curve manifests strong bunching of photons, with a $g^{(2)}$ contrast of ~0.7. To interpret this, we refer firstly to the results for two non-interacting CQDs within the focal volume, which manifest very different behavior (SI section S2). They still show a step-like intensity profile with an off state when both particles are dark, various intermediate states when one or the other emit, and the highest intensity for instances in which both particles are on. The $g^{(2)}$ contrast is ~0.3 for this case, while the theoretical value for two independent single photon emitters is 0.5. The slightly lower value arises from the blinking of the individual particles within the spot (Figure S5). Clearly, the weakly fused CQDM is also different compared with just having two uncoupled CQDs and already manifests coupling effects and different emissive charged and biexciton states leading to its unique optical signatures, compared to the monomers.

The dimer case enables having roughly two types of charged and biexciton states – localized versus delocalized (or segregated). The localized multi-carrier states may be expected to be essentially non-radiative, similar to the CQD monomers, although weak emission may arise due to the slowing down of the non-radiative Auger on account of the larger volume available for the charge carriers. Therefore, such localized states cannot explain the different behavior, pointing out to the contribution from segregated multicarrier states. First, the segregated biexciton, where one electron-hole pair resides on each core, will manifest enhanced $XX$ PLQY due to the significantly slower Auger process (we see that the Auger process is fast in monomers under matched $<N>$ value; Figure S4). Second, the segregated charged trion where the extra electron resides far from the centre hosting the electron-hole pair, will also become emissive. Charged segregated biexciton states may also occur and will be partially emissive.[65-66] All-in-all, these multiple possibilities explain the flickering behaviour between numerous emissive states in the weakly fused CQDMs. The occurrence of the emissive segregated $XX$ state, also explains the significantly reduced $g^{(2)}$ contrast.

Interestingly, the emission behavior of the CQDMs drastically alters at the strong coupling regime. When the neck is filled at the dimer interface, forming a rod-like architecture, the emission trajectory manifests a more step-like behavior, partially regaining the blinking feature as in single CQDs (Figure 2). Along with this, the $g^{(2)}$ contrast is increased back to 0.25, back to within the single photon emitter regime. The inferred drastic reduction of the $XX$ PLQY and reappearance of discernible ON-off states with a small change in the neck diameter in the CQDMs are interesting and non-intuitive. These transformations of the emission characteristics from single emitter in the monomer, to multiple emitter in the weakly fused case, and back to single emitter in strongly fused CQDMs, points to novel charge redistribution within the artificial molecules as a function of coupling strength.

**Fluorescence lifetime and quantum yield analysis for CQDMs in different coupling regimes**

For further tracking the charge redistribution in the CQDMs compared with monomers, we also studied the emission lifetimes and fitted all traces with a common frame of a triexponential decay function (Fig. 2C). The monomer displays a nearly single exponential decay of ~35ns reflecting the radiative rate of the neutral exciton, $X^0$. A small fast component with a lifetime of 1.5ns is also seen, contributed from the short non-radiative lifetime of the charged- $X^-$ or bi-exciton $XX$ states. A middle component with 15 ns lifetime has a very small contribution and assigned to the rare grey states related to some negative trion $X^-$ emission. The $X^0$ emission dominates the total decay profile of the monomer with a ratio of the areas for $XX: X^-: X^0$ of 0.02: 0.05: 1. This conforms to the single photon emitter characteristics of the monomer CQDs, as the emission from the charged- and bi-exciton states is quenched and does not interfere with the neutral emission.

The PLQY of the various charged- and biexcitonic states can also be quantified from the measured data.[61-63] This considers both the excitation, i.e. bi-exciton creation part, and the stability, i.e. the PLQY of the bi-exciton created. Following the poison distribution, the bi-exciton creation probability, $P_{XX} = 1 - e^{-<N>} - <N> e^{-<N>}$, is lower than that of the single exciton, $P_X = 1 - e^{-<N>}$. The calculated ratios of the creation probabilities of the biexciton to single exciton, $\frac{2P_{XX}}{P_X^2}$, is 1.04 for the monomer. Concerning the emission quantum yield of the particular state, this determines the weight of multi-excitons in the detected photon statistics. The $XX$ PLQY $Q_{XX}$ is calculated from the $g^{(2)}$ contrast as,

$$\frac{g_c}{g_s} = \frac{2P_{XX} \cdot Q_{XX}}{P_{X^2} \cdot Q_X} \qquad (1)$$

And found to be less than 10% for the monomers.

$Q_{XX}$ is also related to the decay rates,

$$Q_{XX} = \frac{k_{XX,r}}{(k_{XX,r} + k_{XX,A} + k_{XX,nr,trap})} \qquad (2)$$

Where $k_{XX,r}, k_{XX,A},$ and $k_{XX,nr,trap}$ are the $XX$ radiative rate, Auger rate and the non-radiative rates imposed by trap assisted decay channels, respectively. Thus, the $Q_{XX}$ largely depends on the $XX$

Auger rate, which is rapid for monomers leading to the observed small $g^{(2)}$ contrast correlated with the small relative weight of the fast $XX$ decay component in the lifetime decay.

In case of the weakly fused dimer, the long component related to the neutral $X^0$ manifests the same 35ns lifetime as in the monomer, while the intermediate and the short components are 12 and 2.5 ns, respectively. The ratio of the areas for for $XX^*$: $X^{*-}$: $X^0$ is very different from the monomer though, and they become comparable, 1.6: 2.2: 1. The intermediate lifetime component corresponds with the radiative rate of the segregated trion, while the short component corresponds to the biexciton. Here the asterisk (*) represents an emissive and hence segregated class of the charged- and bi-exciton cases, unlike in the monomer case where the respective states are non-emissive. The calculated ratio of the creation probabilities of the biexciton to single exciton, $\frac{2P_{XX}}{P_X^2}$, is 1.12 for the dimers, reflecting somewhat enhanced biexciton generation compared to monomers owing to the doubled single exciton absorption cross section. In dimers, the interface at the neck region can impart some defects leading to activation of non-radiative channels. However, for weakly fused dimers the emission trajectories do not access an explicit dark state. This is attributed to the larger dimer volume that can accommodate extra charge and to the possibility of its segregation such that it can be accommodated in the second particle in the CQDM, forming an emissive charged exciton state $X^{-*}$. This is evident from the emission of fast intense photons from the photo-excited CQDMs in spite of scarcity of the neutral exciton species. Note that the observed lifetimes of the charged- and bi-exciton components are slightly shortened relative to those from statistical considerations, in which the radiative rate of the multi-carrier states is faster than that of neutral exciton by a factor of 2 in case of trions (~17ns), and by a factor of 4 in the case of biexcitons (~8ns), due to enhanced recombination possibilities.

Following equation 2, the Auger rates determine the $Q_{XX}$ for the biexciton. Had the biexciton formed within one of the cores in a weakly fused CQDM, the strong Auger interaction is expected to lead to a non-emissive state, similar to the monomer case (sub ns) with low $Q_{XX}$. A shared/segregated $XX$ configuration should manifest weaker Auger interaction, resulting in radiative photon pair emission. This leads to higher PLQY of the $XX^*$ and $X^{-*}$ states as a result of reduction in the Auger rate for the segregated states, accompanied by larger occurrence of charging due to inefficient neutralization of the surface traps, altogether leading to the flickering behavior without well-defined intensity states. The calculated PLQY of the $XX^*$ state from the lifetime data is ~50%, close to the value extracted from the $g^{(2)}$ contrast.

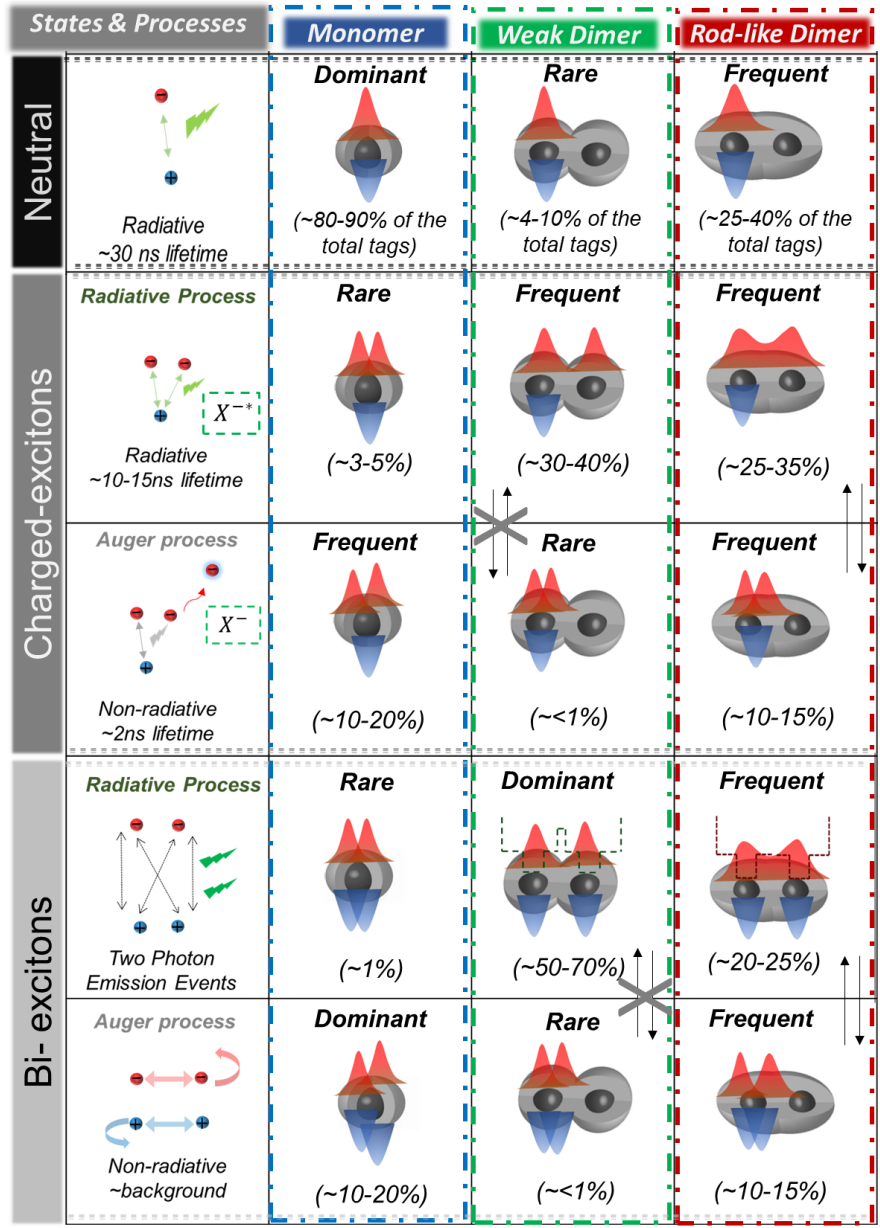

**Scheme 1. Schematic representation of the photoinduced processes in monomers and CQDMs at the different coupling regimes.** In CQDMs, two types of charged- and bi-exciton may form – localized versus segregated. For the segregated cases, the Auger process is reduced among the same charge carriers leading to emissive charged- and bi-excitons. In localized cases, they follow the monomer behaviour where the Auger process is relatively stronger, leading to non-

radiative decay. The neck girth thus plays a crucial role in governing the abundance of the emission and Auger pathways in CQDMs, altering their single-photon emission characteristics.

For the rod-like dimer at the strong coupling regime, the situation reverses. The fitted lifetimes components are 2.5, 11, and 35ns with the ratio of the $XX^*: X^{-*}: X^0$ contributions being 0.2:0.3:1, showing relative increase of the neutral exciton emission component compared with the weakly fused dimer case. The shortest component, with a decay time of 2.8 ns, consists of both radiative and non-radiative contributions from the $XX$ and $X^-$ states, as analyzed by the intensity sliced lifetime data. The effective reduction in the $XX$ PLQY is reflected also in regaining of the single photon emitter characteristics in rod-like dimer with distinct temporally resolved emissive states leading to enhanced $g^{(2)}$ contrast, consistent with an $XX$ QY of 20-25%, close to the value from the lifetime analysis.

**Fluorescence-Lifetime-Intensity Distribution for CQDMs in the different coupling regimes**

The lifetime correlated with emission intensities provides the nature of the exciton dynamics and Auger efficiency within the quantum systems.[61-62,65] Starting with just the intensity distributions, the monomer manifests the well-known bi-modal distribution of emission, which changes over to a rather continuous distribution of the intensities in the dimers, with much stronger emission as discussed above. The weakly fused dimer shows a sharp rise at the low intensity edge, absent of any off states, while in the rod-like dimer, the off state is identified (Figures S6, S7).

The interplay between the excitonic states following charging and discharging events in various coupled and uncoupled quantum systems is best represented in the fluorescence lifetime intensity distributions (FLIDs) in a single frame (Fig. 3B). The FLID of the monomer CQDs reflects the rapid switching between the bright and dark states with an exponential dependence of the intensity versus lifetime, as expected from well passivated core/shell nanocrystals (blue trace).[61-63] A strong correlation between the emission intensities and lifetime can be seen, as the bright (dark) states possess long (short) lifetime. The rare intermediate states seen in the FLID are mostly composed of a linear combination of the two prominent states, as a result of switching events faster than the bin time, further confirming that the emission from the grey states is uncommon.

The weakly fused CQDM manifests uncorrelated behaviour in the FLID data, with a continuous linear change in the intensity and lifetime in contrary to the monomer (Figure 3B, green trace; Figure S6). The FLID distribution is highly confined to short lifetimes, at a wide range of weak to strong intensities. This trace differs considerably from the FLID for a case where there are two uncoupled CQDs within the spot, where the lifetime plateaus while the intensity increases with the number of particles (SI, Figure S5). This indicates that the emission arises from a mixture of emitting configurations with nearly common lifetimes. This causes the photon bunching (as seen in Figure 2) as multiple photons have comparable arrival times to the detector, with an interplay of the fruitfulness of the event, PLQY, and radiative rate of the individual states. The reduction in the Auger rates of the segregated charged- and bi-excitons increases the PLQY of these states, accentuating their dominance in the emission data (Scheme 1).

For the rod-like strongly coupled CQDM, the FLID distribution becomes more broadly distributed on the lifetime axis compared to the weakly fused dimer (Figure 3B, brown trace; Figure S7). Instead of the binary FLID distribution in monomers, a rather broad distribution is resolved, with states possessing intermediate ~9-12ns lifetimes appearing prominently between the long-lived bright-neutral state and the short-lived dimmed state. At this strong coupling limit, charge delocalization is significant, and the interaction among the charges is expected to give rise to richer multicarrier emission. However, this strong interaction reactivates the non-radiative Auger process due to strong delocalization of the electron wavefunctions. The efficient delocalization imparts strong interaction among the carriers and decreases the biexciton PL QY compared with the weakly fused case (Figure S8).

The statistical distributions of the emission intensity, and lifetime of ~35 single particles from each sets are presented in Figure S9-S10. A much defined distribution in each set is found, Figure 3B being a representative of the statistics. The top emission intensity values of both the weakly-fused and rod-like dimers remain ~2000-2500 counts/50 ms, a more than twice value compared to monomers (Figure S9), where the lifetime from these top intensity values are different in two coupling regimes (Figure S10). The values being ~8-12 ns in the weak- coupling regime dominated by emissive charged- and bi-exciton states, while in the strong coupling regime the values reach ~25-30ns akin to emission from neutral exciton state in monomer CQDs.

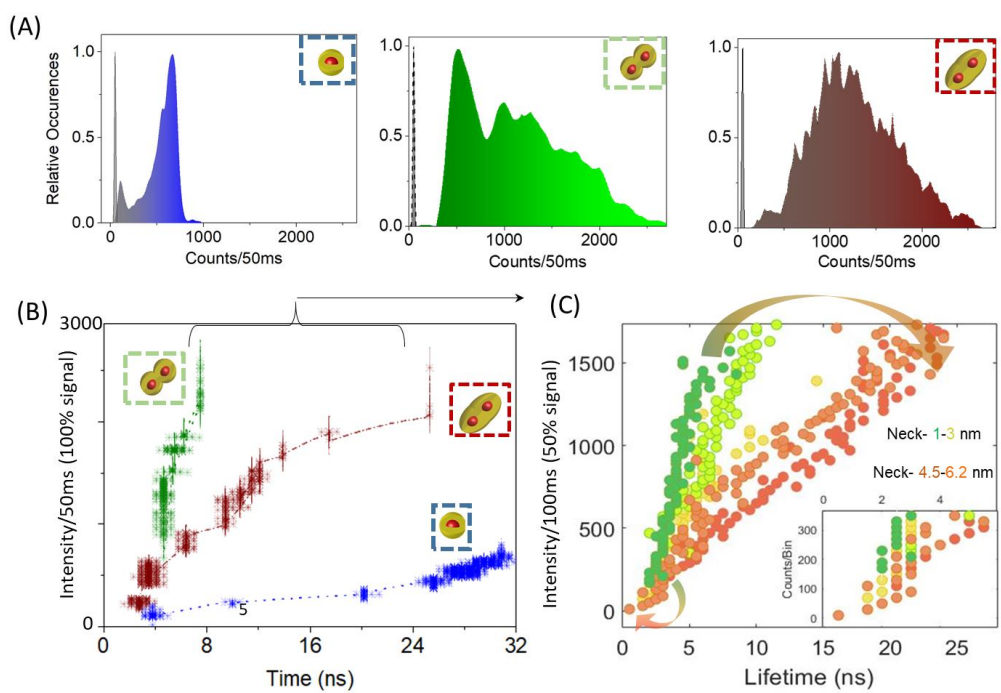

**Figure 3. Fluorescence intensity – lifetime correlations as a function of neck girth** (A) Fluorescence intensity distribution of single monomer, weakly-fused dimer, and rodlike dimer; acquisition time 120s, bin time 50ms. The sharp grey curve represents the background counts. (B) Fluorescence-Lifetime-Intensity-Distribution (FLID) plots for the same particles. (C) FLID diagram of dimers with varying neck girth ranging from weak-to-strong coupling. Upon neck filling alteration towards longer lived neutral exciton emission is observed. Inset shows the zoomed plot at early values, demonstrating the re-generation of the low intensity-short lifetime off states upon neck filling.

**Facet-limited versus beyond-facet attachment dictates the coupling regime**

Recapping, in both types of dimers, three distinct states are observed with lifetime of 30-35ns ($X^0$), 9-12ns ($X^{-*}$) and 2-3ns ($XX, X^-$) with descending order of emission intensities. The relative population of the three states are different in the weakly fused and rodlike dimer cases. We further examine this transition by studying the intermediate fusion regime between these two limits, where through suitable synthesis control, we achieved a series of CQDMs samples with systematically varying neck girth and thus corresponding intermediate electronic coupling strengths.[26] Figure 3C presents the FLID analysis of CQDMs with increasing neck girth (from green to red points). The green to light yellow points refer to the CQDMs with neck diameter of ~1-3nm (smaller than the fused facet size) and from orange to red points refer to the CQDMs with neck diameter of ~4.5-6.2nm (larger than the fused facet size). The single particle emission intensities measured under identical conditions are fairly similar in all the CQDMs, but a systematic change is observed in terms of the lifetimes where the thicker-neck CQDMs manifest longer decays (Figure S8-S10).

In particular, the FLID data is distributed into two groups. In the first group, with narrow neck diameters, the change in the FLID is linear and only faster decays dominate. To interpret this, recall that charging is a common phenomenon in photo-excited CQDs; once the particle gets charged via an Auger process or by hole trapping, it requires energy for the detrapping.[62,67] The charging/trapping and the discharging/de-trapping in CdSe/CdS CQDs is generally attributed to "hot-carrier blinking", leading to their fluorescence blinking phenomena.[61-62] In this mechanism, there are incidences in which the hole traps at the surface leaving behind a negatively charged CQD. Upon excitation, the presence of the three carriers leads to an Auger process in which the electron-hole pair transfers its energy to the electron that is becoming highly energetic ("hot") and can thus also enable the detrapping process (Scheme 1). In the regime of narrow neck girth, below the facet size, the observed squeezing of the FLID diagram to low lifetime values indicates that the charged states are dominant and emissive. Here, unlike in monomers, the Auger process is weak due to weak hybridization and the neutralization process is also inefficient (inset of Figure 3C; Scheme 1). The outcome is that the CQDM is often charged, yet emissive due to the segregation of the extra carrier to the other dot. Similarly, the $XX$, or $XX^-$ states are also emissive due to the carrier segregation.

In the second group, beyond the facet limited attachment ("super-necking"), the lifetime data become stretched towards the $X^0$ state (~30-35ns) with a breakpoint at the charged regime (~400-700 counts/bin; ~7-10ns lifetime). Here, hybridization is strong and correspondingly the Auger processes are more efficient. This facilitates faster neutralization of the CQDMs, as the hot electron recombines with a trapped hole leading back to the neutral-on state.[59] The emergence of the Auger induced hot electron is also evidenced at the lower values of the FLID diagram (inset of Figure 3C), where the off states appear for neck girth beyond facet size.

It is therefore interesting to summarize that while the monomer CQD is a very good single photon emitter, in CQDMs, for the "below facet" attachment regime we observe tendency towards two photon emission. The single photon emitter characteristics are regained upon reaching the "beyond facet" attachment regime (Figure S8).

**Spectral signatures of the charged- and bi-exciton states in CQDMs**

We next correlate the time-tagged data with the spectral features of the CQDMs in the different coupling regimes. For this, we have divided the emission signal by a suitable beam splitter into two parts, 50% to the APDs, and 50% to the spectrograph coupled with EMCCD. Simultaneous collection of the time-tagged-time resolved (TTTR) data, along with spectral information, further ascertain the analysis by the FLID plot and the proposed mechanisms in Scheme 1. Figure 4 shows the emission intensity level dependent $g^{(2)}$ curves, and accumulated emission spectra for weakly coupled (upper panels A-D), and rod-like CQDMs (lower panels E-H). The red, black and blue traces of $g^{(2)}$, spectrum, and lifetime curves represent the data from the high-, intermediate- and low- intensity levels, respectively (Figure 4 A, E). We find that for the weakly fused dimers, although no clear difference between the states is observed, when the second-order correlation function is constructed from the top-most intensity tags, ~43% of double photon events are seen (red, Figure 4B), arising from the influence of the charged- and biexciton events rapidly switching throughout the measurement period. The bottom most emissive tags, however, provide near unity (0.85) value of the $g^{(2)}$, indicating that it is dominated by *XX* emission, (blue, Figure 4C). The corresponding spectra are quite broad, due to flickering between multiple emissive states, but show slight blue-shift for the predominately bi-exciton tags, indicative of a repulsive segregated biexciton (Scheme1).[52,68] The biexciton emission follows a cascade pathway in a three-level system: XX→X+hv followed by X→ ground state+hv.[53,69] In the weak-coupling regime, the electron wavefunctions are mostly localized on each dot, reducing the overlap between the electron in dot 1 and hole in dot 2. This indirect attractive Coulomb force is weaker than the Coulomb attraction of electron and hole in each dot. However, there is still coupling between the electrons, and the interactions between all charges are possible. Additionally, repulsion between the similar charges enables Auger interaction, which reduces the PLQY of the XX to ~50%. The high $g^{(2)}$ contrast at the linear excitation regime($< N > < 0.2$) indicate the potential of the "connected-but-confined" CQDMs for efficient *XX* generation in photonic applications.[70-72]

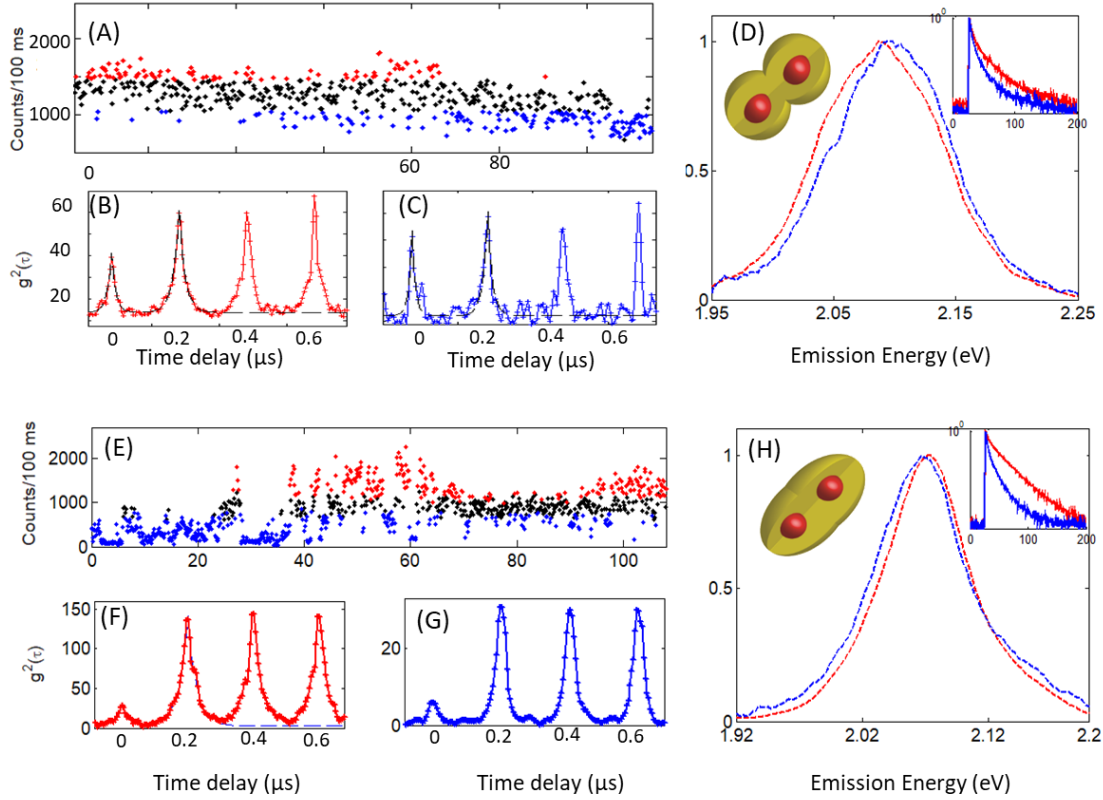

**Figure 4. Spectral feature of the different states in various artificial constructs, derived from spectral-TTTR correlation.** (A-D) TTTR data for a weakly fused dimer. (A) Time trace intensity data distributed into three intensity regions from high (red) to low (blue) (B,C) Intensity dependent g(2) curves manifesting emissive bi-exciton contribution. (D) Emission intensity dependent spectra and lifetime (inset). (E-H) TTTR data for a strongly fused dimer with similar analysis as for (A-D). Much weaker biexciton emission leads to regained single photon emitter characteristics. In case of rod-like CQDMs, only a moderate enhancement of the g(2) contrast is seen for the upper (~0.14) to the lower intensity tags (~0.25), indicating much fewer bi-excitons emission, assigned to the presence of efficient non-radiative Auger decay channels (Scheme 1). This is also reflected in the spectra, as the lower intensity tags predominately exhibit a red-shift in the emission maxima (attributed to a negative trion shift), with a small blue-shifted hump, compared to its own top-intensity tags (Figure 4H). In such "connected-and-delocalized" CQDMs, the *XX* emission is low and the emission is dominated by charged followed by neutral excitons due to reactivation of the Auger decay, unavailable for the weakly-fused, "connected-but-confined" CQDMs.

**Correlating coupling strength with spectral characteristics of CQDMs**

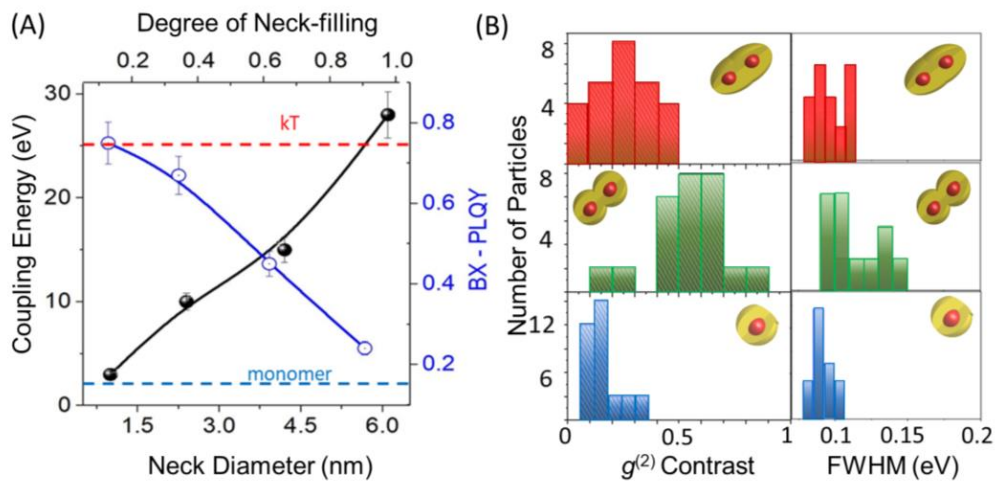

**Figure 5. Emitter characteristics versus coupling strength for CQMs** (A) Plot of Coupling strength (black) and bi-exciton PLQY (blue) as a function of the neck girth. The blue dotted line represent the monomer case as a guide to eye. (B) The statistical data (30 particles) for the g(2) contrast (left), and spectral width (right) for monomers (blue), weakly fused CQDMs (green) and rodlike CQDMs (red). In weakly coupled CQDMs high g(2) contrast is accompanied by broad emission reflecting the emissive charged and biexciton states. In strongly coupled CQDMs both are regained to lower values, closer to those of the monomers.

We find that alongside the hybridization, the photon correlation characteristics in CQDMs are also tuned systematically upon changing the neck barrier at the dimer interface. Figure 5A shows the dependence of the hybridization energy extracted from the red shift in the emission ("coupling strength"), together with the extracted $XX - PLQY$, versus the neck girth. Weakly fused CQDMs manifest small coupling energy and bunching of photons on account of emissive charged and bi-exciton states. The strong coupling regime with larger hybridization energy, regains partially the photon purity as reflected by the decreased antibunching contrast. (Figure 5B) The spectral widths at room temperature follow the similar trend. A narrow spectrum of the monomer of FWHM ~70-85 meV was obtained upon averaging ~30 single particles, indicating exclusive neutral exciton emission while considering size distribution and spectral diffusion at room temperature (Figure 5B). In the weakly fused CQDMs, the FWHM of the single particle spectra broaden to ~150meV reflecting the collective emission from all the states with varying energies. However, at the strong coupling regime the situation reverses, as the electron delocalization modifies the Auger and detrapping processes. The spectral width of the single particles narrows back to ~90-110 meV, only slightly broader than the monomers, as expected from emission by the neutral exciton and the stable trion states, which differ in their respective PLQY (Figure S11).

**Conclusion and Outlook**

The single particle spectroscopic analyses for the epitaxially connected CQD dimer molecules with various coupling regimes were presented, revealing that the extent of coupling can tune the emitter's characteristics. Along with the hybridization energy, the necking parameter controls the Auger interaction, relative abundance, and the stability of the possible emissive states. The varying radiative and non-radiative rates depending on the neck girth affect the relative abundance of the neutral single exciton and biexciton states thus impacting the observed photon statistics.[73-75] Single CQD monomers exhibit mostly neutral exciton emission, $X^0$, due to rapid charge neutralization. The weakly coupled CQDMs manifest relatively high *XX PLQY* and are highly charged. As the coupling strength rises with increasing neck girth, inclination towards neutral exciton emission grows gradually, reducing the quantum yield of the two photon emission events. This switching of emission characteristics from weakly-to-strongly coupled CQDMs, are analogous to transitioning between "ionic-type" to "covalent-type" bonding in naturally occurring molecules.[76] Already at the weak coupling regime, the excitons and charges interact strongly. At the strong coupling limit, the CQDMs behave as quasi- single photon emitters with higher intensity compared to the monomers under similar excitation conditions. As a result of the removal of the neck barrier, such CQDMs manifest distinct neutral and charged states. Non-radiative Auger processes are activated, when the electronic coupling energy approaches to the thermal energy.[77-79] Controlling the neck girth in CQDMs offers a handle to manipulate the coupling strength and also the photon statistics and emissive states. Thus neck-engineered CQDMs can be further considered for future quantum technologies applications.[20, 80-88]

**EXPERIMENTAL PROCEDURES**

**CQDM synthesis and sample preparation for single particle study**

CQDMs with various neck diameters were synthesized as per our recently established "neck-filling via intra-particle ripening mechanism".[26] Chemically linked CQD dimers are prepared on template silica particles, followed by release of the linked dimers in solution. Thermo-chemical fusion reaction, leading to neck formation at the interface of CQD dimers.[22-26] The diameter of the neck is governed by the proper tuning of the fusion-reaction parameters including the temperature and free surface ligand concentration.[26] Starting from a non-fused chemically linked CQD dimer, weakly fused CQDMs (Neck diameter < 3nm) was achieved with addition of 10% free ligand (Oleic acid + Oleyl-amine) at $200^0C$ for 20h. For a rod-like CQDM sample preparation (Neck diameter > 4.5nm), the added ligand concentration was reduced to ~1% at elevated reaction temperature ($240^0C$) for 20h. Intermediate neck diameter samples were also obtained by fine tuning of the reaction temperature and ligand concentration. Next, we employed rigorous size selective precipitation technique to enhance the dimer concentration in a particular sample set, monitored with TEM and ensemble optical measurements[22,26]. We obtained four different sets of fused dimer with different neck diameters, exhibiting distinct and systematic optical properties. Two sets of weakly fused dimer (with facet-limited attachment) and two sets of strongly fused dimer (beyond facet attachment) were obtained, and studied.

**Single-Particle-Spectroscopy of CQDs and CQDMs**

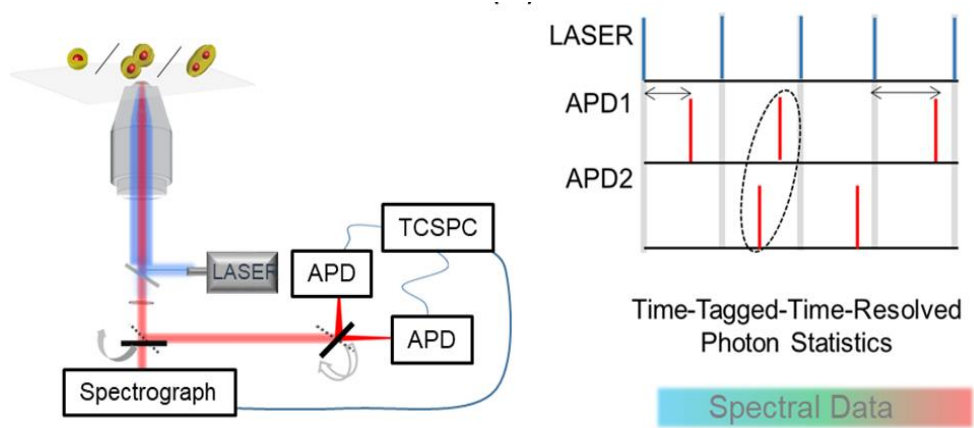

**Scheme 2: Single particle measurement setup.** The PL emission from single particles is analyzed with time-tagged-time-resolved (TTTR) photon statistics along with simultaneous spectral information (APD – Avalanche photodiode, TCSPC – time correlated single photon counting).

For single-particle fluorescence measurements the cleaned and diluted CQD and CQDMs are spin coated on glass coverslip (thickness ~150μm). Single particle spectroscopy was performed with an inverted optical microscope with epi-luminescence configuration. The particles were excited with a 405nm pulsed laser (~50ps pulse duration, repetition rate- 5MHz) passed via a 450nm long-pass dichroic mirror, through 100X oil immersion objective with 1.4 NA. The emission was collected and sent to detection through a 500nm long pass filter to remove any back reflection from the laser source. In the detection part, we have analysed the emission signal with two Avalanche Photodiode (APDs) at a Hanbury-Brown-Twiss (HBT) geometry, and with a spectrograph coupled with Electron Multiplying Charge Coupled Device (EMCCD), individually or simultaneously. All the detectors are connected with a common Time-Tagger to synchronize the signal from the detectors, and for time-tagged-time-resolved analysis. The data represented in Figures 2 and 3B are obtained from 100% APD signal (bin time 50ms), For Figures 3C and 4, the emission signal was divided with a 50/50 beam-splitter to the APDs and the spectrograph to acquire the signal simultaneously (bin time 100ms). The emission signal in the detectors was analysed with suitable MATLAB codes. The data represented here was verified with different batches of CQDMs and are consistent with the presented concept. In each cases, the monomer out of the dimer, obtained from the size selective precipitation, was studied as a standard to verify any inherent change in the NCs following the neck-filling synthesis procedures.

All the experiments are performed under low photo-excitation intensity keeping the number of exciton produced per pulse $<N> = \sigma \times J = 0.1$ for monomers ($\sigma$ is the absorption cross-section, and $J$ is the excitation flux density per unit area). We chose this value of $<N>$ to generate emission signal with high signal-to-noise ratio in the linear excitation regime. Moreover, the extraction of $XX$ PLQY following HBT measurements is valid in this modest excitation condition.[54] We observe that the emission contributions in TTTR data from the neutral- ($X^0$), charged- ($X^-$), and bi-exciton ($XX$) vary systematically following the poison statistics and photo-charging. Any higher excitation lead to disappearance of the neutral-exciton ($X^0$) contribution in case of weakly fused CQDM, the characteristics of which remain a metric for the determination of lifetime, PLQY of the charged- and bi-exciton in a particular system under study.

AUTHOR INFORMATION

**Corresponding Author**

* Correspondence and requests for materials should be addressed to U.B. (uri.banin@mail.huji.ac.il)

**Present Addresses**

† Present address: The Center for Molecular Imaging and Nuclear Medicine, State Key Laboratory of Radiation Medicine and Protection, School for Radiological and Interdisciplinary Sciences

(RAD-X) and Collaborative Innovation Center of Radiological Medicine of Jiangsu Higher Education Institutions, Soochow University, Suzhou 215123, China.

**Author Contributions**

S.K. and U.B. conceived the idea. S.K. performed the optical experiments, analyzed, and represented data with input from Y.O. and E.S.. J.C and A.L synthesized and purified the CQDs and CQDMs, S.K. took part in size selective separation of dimers. S.K., U.B. Y.E.P. interpreted the results to declutter the processes, with inputs from all other authors. S.K. and U.B. co-wrote the manuscript with contribution from all authors. U.B. supervised the work.

**Acknowledgements**

The research leading to these results has received financial support from the European Research Council (ERC) under the European Union's Horizon 2020 research and innovation programme (grant agreement No [741767], advanced investigator grant CoupledNC). S.K. and J.C acknowledge the support from the Planning and Budgeting Committee of the higher board of education in Israel through a fellowship. Y.E.P. acknowledges support by the Ministry of Science and Technology & the National Foundation for Applied and Engineering Sciences, Israel. U.B. thanks the Alfred & Erica Larisch memorial chair. We thank Professor Eran Rabani for stimulating discussions. We thank Dr. Sergei Remmenik at the HUJI Nanocenter for assistance in structural characterization of the CQDMs.

**DECLARATION OF INTEREST**

The authors declare no competing interests.

8. Ekimov A. I., Efros A. L., and Onushchenko A. A. (1985). Quantum size effect in semiconductor microcrystals. *Solid State Commun. 56*, 921–924.
9. Brus L. (1986). Electronic wave functions in semiconductor clusters: Experiment and theory. *J. Phys. Chem. 90*, 2555–2560.
10. Fojtík, A., Weller, H., Koch, U., and Henglein, A. (1984). Photo-Chemistry of Colloidal Metal Sulfides 8. Photo-Physics of Extremely Small CdS Particles: Q-State CdS and Magic Agglomeration Numbers. *Berichte der Bunsengesellschaft für physikalische Chemie*, *88*, 969-977.
11. Banin, U., Cao, Y., Katz, D., and Millo, O. (1999). Identification of atomic-like electronic states in indium arsenide nanocrystal quantum dots. *Nature 400*, 542-544.
12. Panfil, Y.E., Oded, M., and Banin, U. (2018). Colloidal quantum nanostructures: emerging materials for display applications. *Angewandte Chemie International Edition 57*, 4274-4295.
13. Supran G. J., Shirasaki Y., Song K. W., Caruge J.-M., Kazlas P. T., Coe-Sullivan S., Andrew T. L., Bawendi M. G., and Bulović V. (2013). QLEDs for displays and solid-state lighting. *MRS Bull. 38*, 703–711.
14. Chen, O., Zhao, J., Chauhan, V. P., Cui, J., Wong, C., Harris, D. K., Wei, H., Han, H. S., Fukumura, D., Jain, R. K., and Bawendi, M. G. (2013). Compact high-quality CdSe–CdS core–shell nanocrystals with narrow emission linewidths and suppressed blinking. *Nature materials 12*, 445-451.
15. Ji, B., Koley, S., Slobodkin, I., Remennik, S., and Banin, U. (2020). ZnSe/ZnS core/shell quantum dots with superior optical properties through thermodynamic shell growth. *Nano letters 20*, 2387-2395.
16. Won, Y. H., Cho, O., Kim, T., Chung, D. Y., Kim, T., Chung, H., Jang, H., Lee, J., Kim, D., and Jang, E. (2019). Highly efficient and stable InP/ZnSe/ZnS quantum dot light-emitting diodes. *Nature 575*, 634-638.
17. Kim, T., Kim, K. H., Kim, S., Choi, S. M., Jang, H., Seo, H. K., Lee, H., Chung, D. Y. and Jang, E. (2020). Efficient and stable blue quantum dot light-emitting diode. *Nature 586*, 385-389.
18. Chang, Y., Kim, D. H., Zhou, K., Jeong, M. G., Park, S., Kwon, Y., Hong, T. M., Noh, J. and Ryu, S. H. (2021). Improved resolution in single-molecule localization microscopy using QD-PAINT. *Experimental & molecular medicine 53*, 384-392.
19. Park, Y. S., Roh, J., Diroll, B. T., Schaller, R. D., and Klimov, V. I. (2021). Colloidal quantum dot lasers. *Nature Reviews Materials 6*, 382-401.
20. Kagan, C. R., Bassett, L. C., Murray, C. B., and Thompson, S. M. (2020). Colloidal quantum dots as platforms for quantum information science. *Chemical Reviews*, *121*, 3186-3233.
21. Vajner, D. A., Rickert, L., Gao, T., Kaymazlar, K., and Heindel, T. (2021). Quantum Communication Using Semiconductor Quantum Dots. *arXiv preprint arXiv:2108.13877*.
22. Cui, J., Panfil, Y. E., Koley, S., Shamalia, D., Waiskopf, N., Remennik, S., Popov, I., Oded, M., and Banin, U. (2019). Colloidal quantum dot molecules manifesting quantum coupling at room temperature. *Nature communications*, *10*, 1-10.
23. Panfil, Y.E., Shamalia, D., Cui, J., Koley, S., and Banin, U. (2019). Electronic coupling in colloidal quantum dot molecules; the case of CdSe/CdS core/shell homodimers. *J. Chem. Phys.*, *151*, 224501.
24. Koley, S., Cui, J., Panfil, Y. E., and Banin, U. (2021). Coupled colloidal quantum dot molecules. *Accounts of Chemical Research*, *54*, 1178-1188.
23

**Graphical Abstract:**

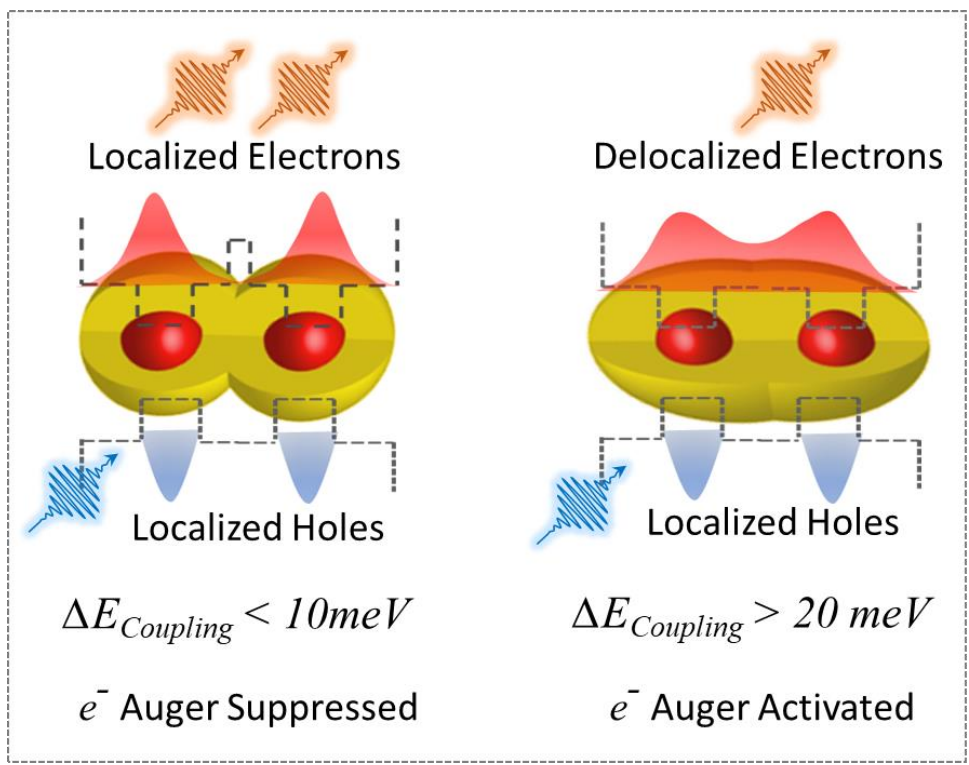

**Highlights**

- Two distinct coupling regimes are identified in "Artificial CQD Molecules"
- Neck width acts as a flexible handle to tailor the emitter characteristics.
- Hybridization strength dictates the charging and neutralization of CQDMs.
- Novel Auger process involving two dots is demonstrated.

# Photon Correlations in Colloidal Quantum Dot Molecules Controlled by the Neck Barrier

*Somnath Koley[1], Jiabin Cui[1,†], Yossef. E. Panfil[1], Yonatan Ossia[1], Adar Levi[1], Einav Scharf[1], Lior Virbitsky[1], and Uri Banin[1]\**

[1]Institute of Chemistry and the Center for Nanoscience and Nanotechnology, The Hebrew University of Jerusalem, Jerusalem 91904, Israel.

Table of Contents

## Section S1. Structure, and Single Particle Emission Characteristics of Monomer:

The CQD monomers (CdSe/CdS core/shell nanocrystals) possess very homogeneous size and shape distribution (Figure S1A) as also reflected on the single particle optical data.

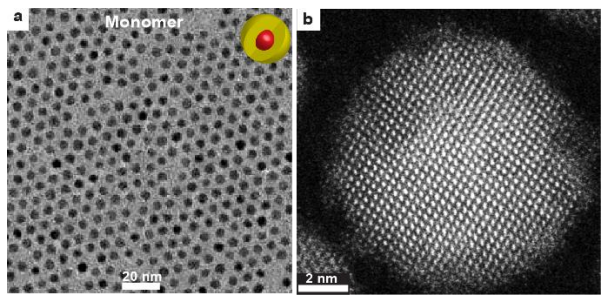

**Figure S1** Representative TEM images of the studied CQD monomers.

The chosen monomer possess quasi type-II band structure and possess long exciton lifetime (~30-35ns). In the blinking trace (Figure S1 B) we observe ~80-90% ON state arising from neutral exciton (blue area in S1 B). Rest of the population appears from the radiative trion (green area) and non-radiative charged- and bi-exciton (red area) as a result of efficient Auger recombination. Figure S1 C denotes different exciton configuration with their corresponding radiative rates in our monomer CQD. Figure S1D represents the rare radiative trion (likely negative trion).

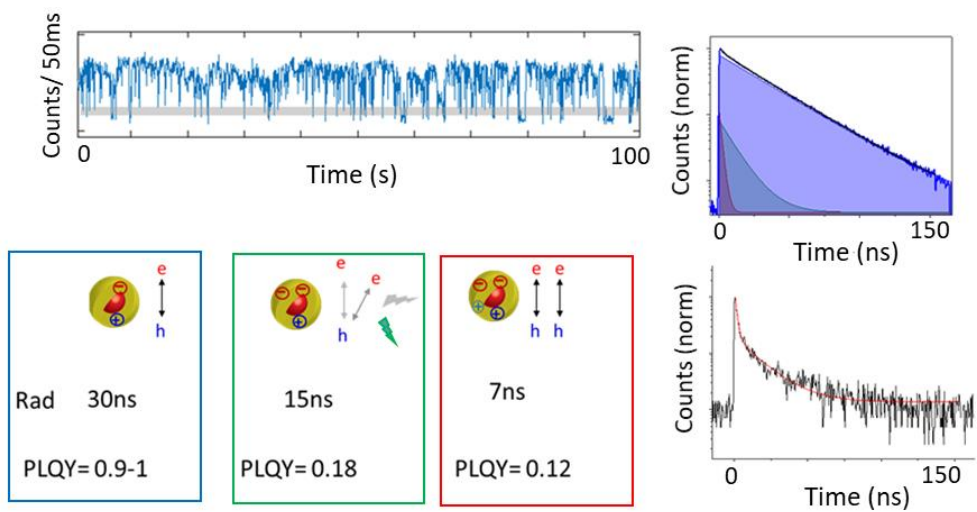

**Figure S2.** Monomer studies on emissive trion in the emission time trace. The emission from single CQD monomer mostly contains ON and off state (non-radiative trion and bi-excitons). Only a few tags were found with only a fraction of radiative trion with ~15ns lifetime.

**Single-particle characteristics of various Monomers:**

**(Separated batch from size-selective precipitation)**

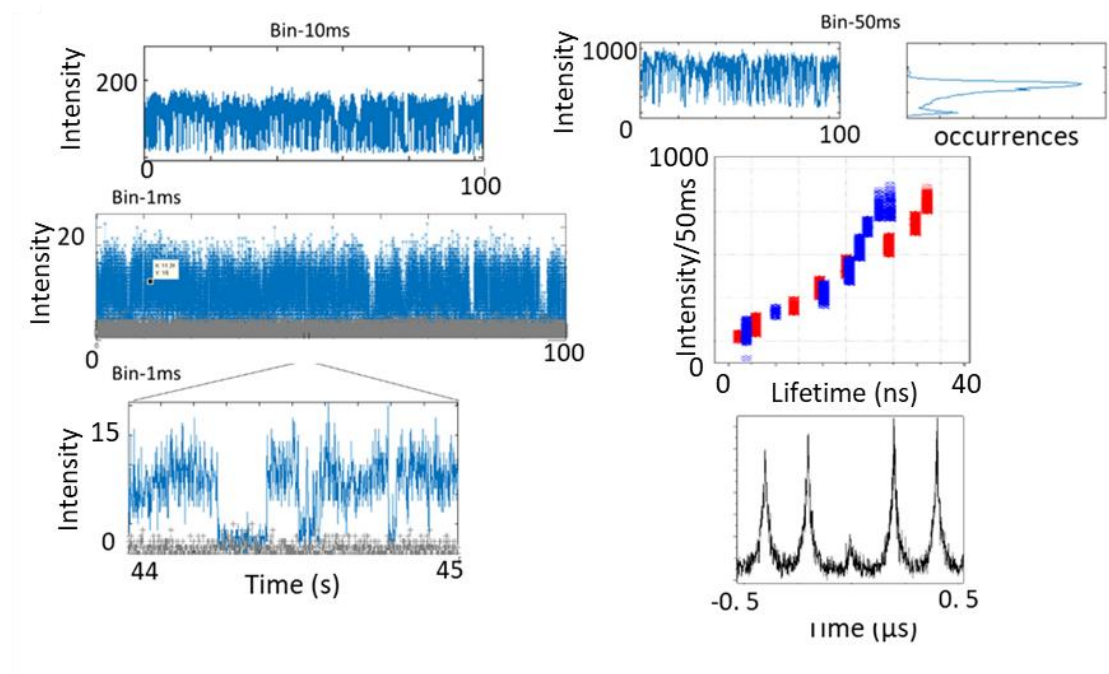

**Figure S3.** We explored the various monomers via TTTR photon correlation experiments at identical conditions. We observe a small deviation among the majority of the particles, and follows mostly "Type-I" Auger blinking (ref. S1).

**Single-Particle Characteristics of Monomer at higher power:**

**(To match the <N> value with dimer)**

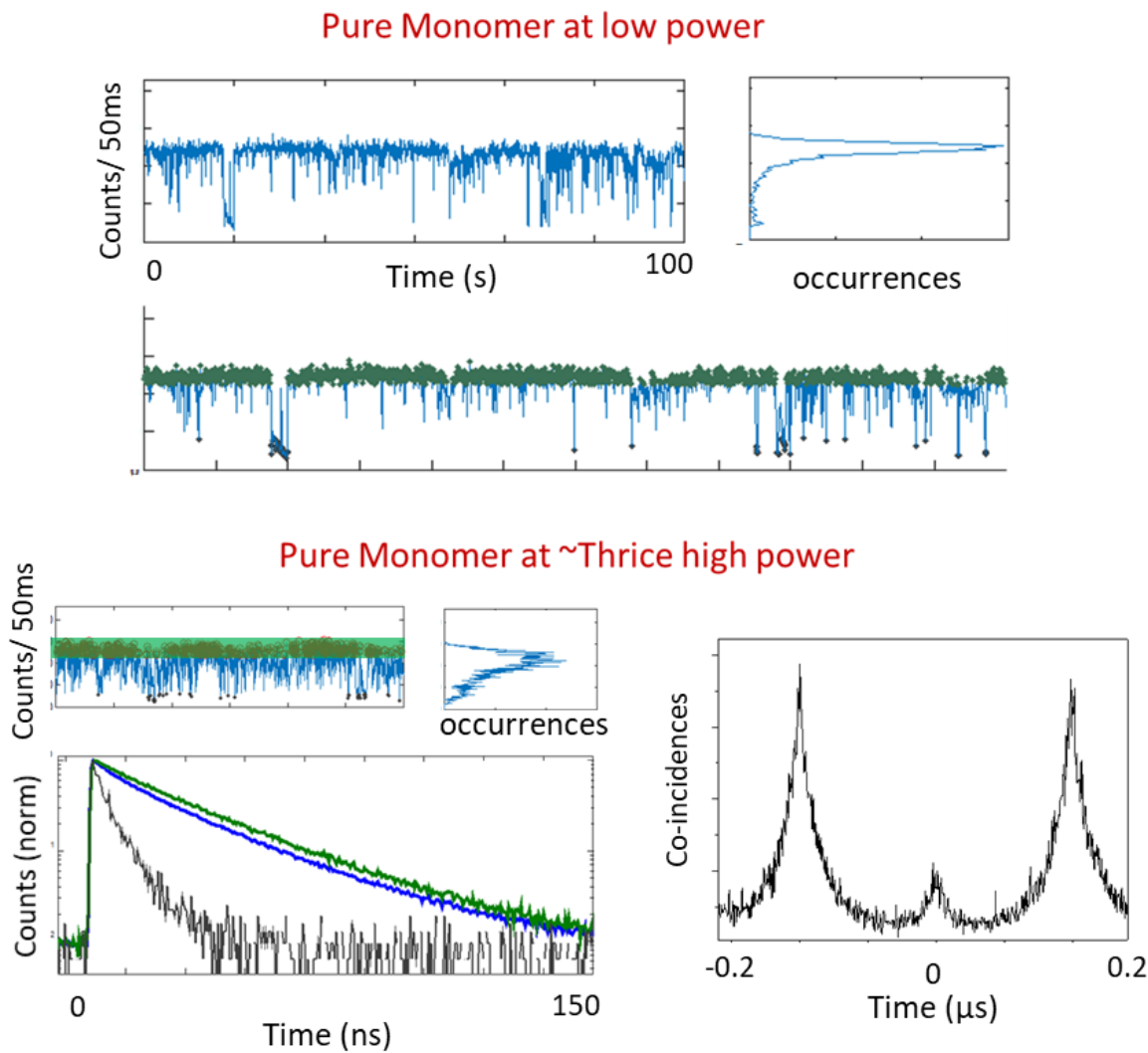

**Figure S4**. Here the pure monomers were excited at higher power to meet the discrepancy between the numbers of exciton produced per pulse (<N>), in monomer and dimer. We see an enhanced blinking, and more contribution of radiative multi-carrier states on the "off" regime, at higher excitation of the same particle. But the bi-exciton PLQY doesn't rise significantly, because of the low PLQY of the state at high power.

## Section S2. Optical Characteristics of the Uncoupled CQD Aggregates:

**(Forming a single diffraction limited spot with narrow spectrum)**

We studied the uncoupled dimer/trimer formed during the spin coating of monomers under the same focal volume. Illumination causes formation of single diffraction limited spot, so the two/three particles must be at proximity of each other. A step-like intensity time trace can be seen whether one or more particles are ON. The theoretical value of 0.5 should be obtained for two particle case, considering both of the CQDs are ON throughout the measurement. But each particle have their own blinking process, hence we obtain a much lower value of ~0.2 in case of two particles aggregate and ~0.3 for three particles aggregate. But almost no coupling is observed as seen by step like intensity profile. In some cases weak FRET interaction take part and affect the top intensity value and lifetime extracted from each step.

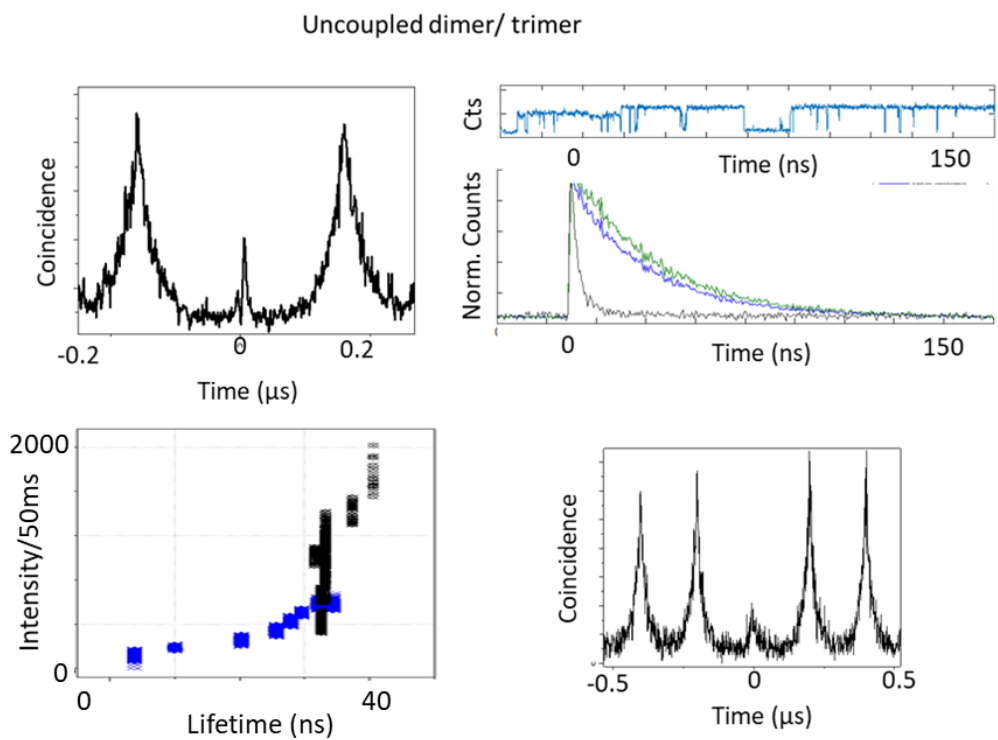

**Figure S5.** The uncoupled dimer and trimer cases, as explained in Page 7 and 11 of the main manuscript. (A) Intensity profile, (B) Sliced fluorescence lifetime decay, and (C) second order correlation function, from the two CQD aggregate case. (D) FLID profile for the three particle case (black), along with the monomer (blue). The intensity is presented at the x-axis for the clarity of the saturation behavior of the lifetime as a function of intensity. (E) Second order correlation function from the emission of three particle aggregate case.

## Section S3. Optical Properties of CQDM at weak coupling regime:
**(An extreme case in the weak-coupling regime)**

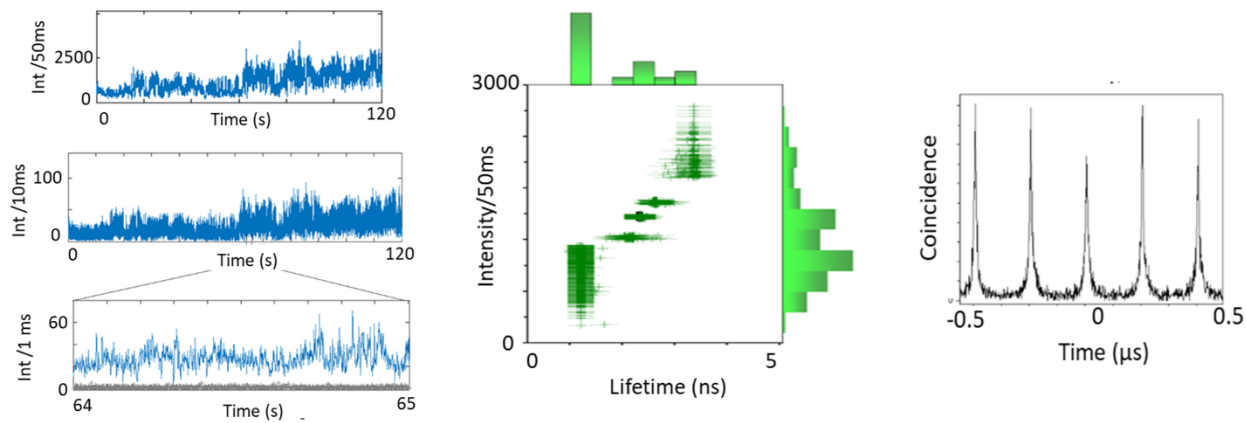

**Figure S6.** Intensity time trace, FLID and the second order photon correlation for a single weakly fused dimer. In case of weakly fused dimer, a very confined FLID is observed ranging ~500-2700 counts/50ms in the intensity axis, and ~3.5-5ns in the lifetime axis. This represent an extreme case of the weak-coupling limit (assigned to ~1nm neck diameter).

## Section S4. Optical Properties of CQDM at strong coupling regime:

**(A near extreme case in the strong-coupling regime)**

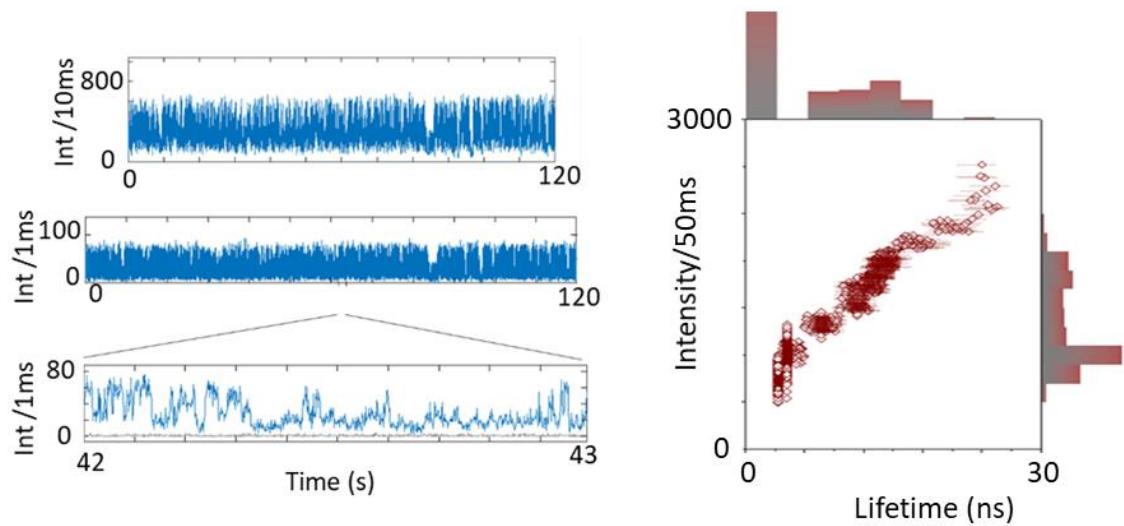

**Figure S7.** Intensity time trace, FLID and the second order photon correlation for a single rod-like dimer. In case of rod-like dimer, a rather stretched FLID is observed ranging ~50-2700 counts/50ms in the intensity axis, and ~1.5-23 ns in the lifetime axis. This represent a near extreme case of the strong-coupling limit (~6nm neck diameter).

**Section S5. Variation among strongly fused dimers:**

Various types of particles within the strong coupling limit. Depending on the extent of fusion, the CQDMs exhibit different extent of electronic coupling. They also exhibit different single particle photo-physics as well. The upper panel represents a CQDM from rod-like dimer batch (neck diameter~6nm), where they are found exclusively in TEM measurements, and the lower panel represents a CQDM from the strongly fused batch (neck diameter~4.5nm). In the former case the fast lifetime is relatively less abundant (red) and has more non-radiative contribution, more also evident from the time traces and FLID profile than the latter case. This affect the g(2) contrast accordingly, as the radiative bi-exciton population decreases with more electronic coupling, and hence Auger activation.

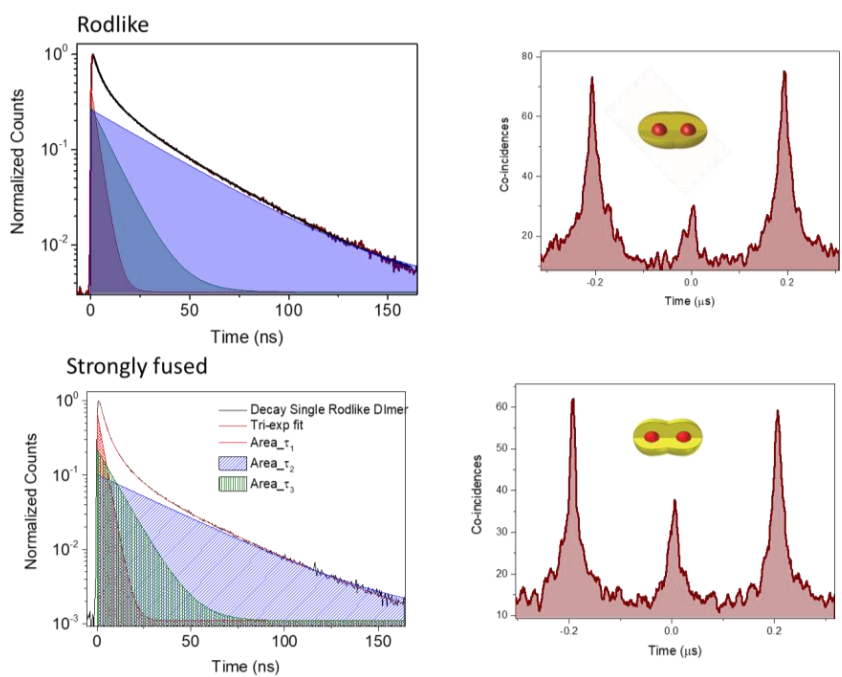

**Figure S8.** Lifetime and $g^{(2)}$ correlation for two particles at the strong coupling regime. Contribution of the neutral single exciton (blue shade) in the lifetime decay is higher, compared to the short bi-exciton component (red shade) in the top case, exhibiting $g^{(2)}$ contrast of ~0.2. A signature of the neck-filled rod-like dimer. While the lower case demonstrate a property in between the weakly fused a rod-like case. Here the neutral exciton contribution (blue shade) is fairly lower than the former case, due to higher PLQY of the bi-exciton state leading to $g^{(2)}$ contrast of ~0.4.

**Section S6. Statistical distributions of emission intensities in CQDs and CQDMs:**

We compare the statistical distribution of emission intensities from a group (~35) of single CQDs, weak-CQDMs, and strong-CQDMs. We see the peak intensities (from intensity distribution plots for each particles) do not vary too much, but the top intensities vary due to stronger absorption in case of CQDMs compared to CQDs.

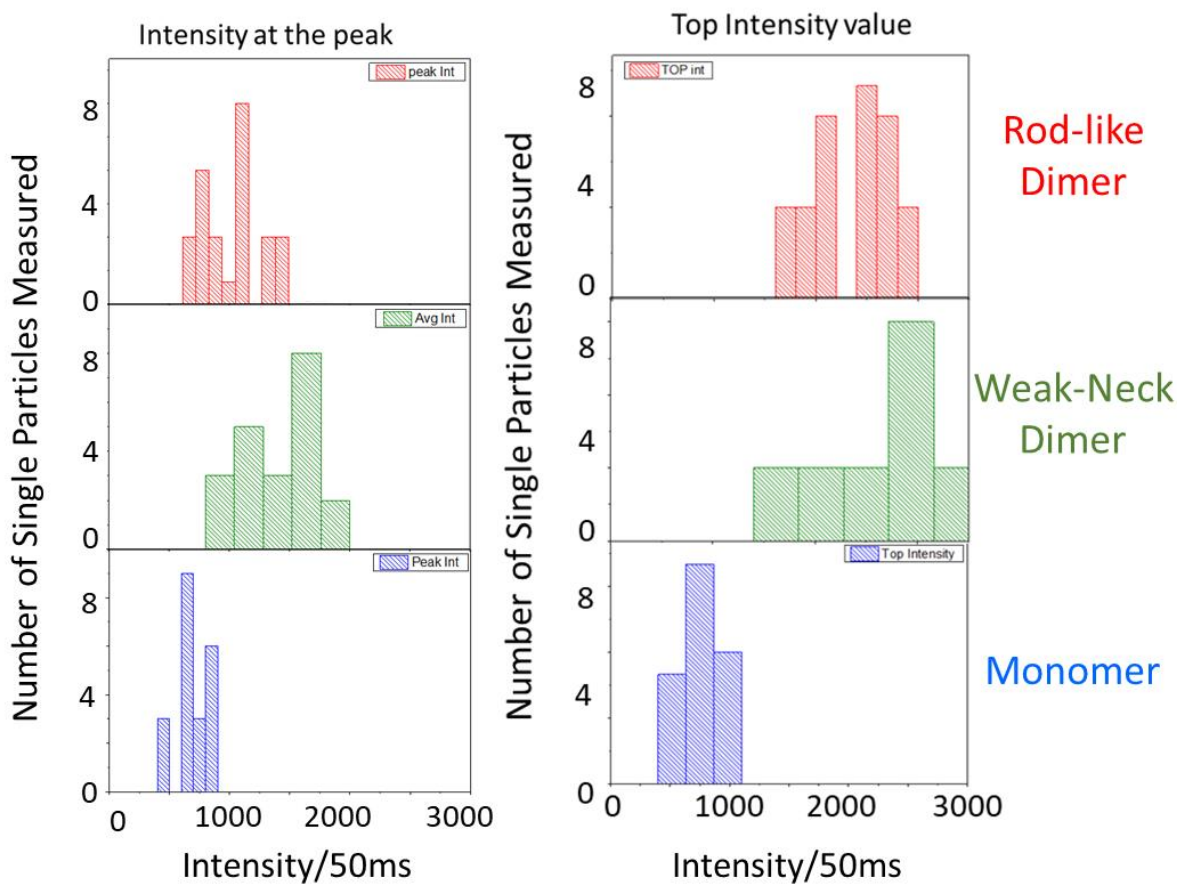

**Figure S9.** Statistical distribution of the single particle emission intensities of various particle (~35-40 particles in each set). Histogram of the peak intensities from the overall emission tags of various particles are shown on the left. On the right side, intensity value of the top 10% tags from various particles are presented.

**Section S7. Statistical distributions of average fluorescence lifetimes in CQDs and CQDMs:**

The statistical distribution of the emission intensities, the emission lifetime exhibit distinct distribution in various constructs. The average lifetime exhibit only a marginal change between the weak-CQDMs and strong-CQDMs. Whereas the top- intensity lifetimes exhibit regeneration of long exciton lifetime in strongly-fused CQDMs.

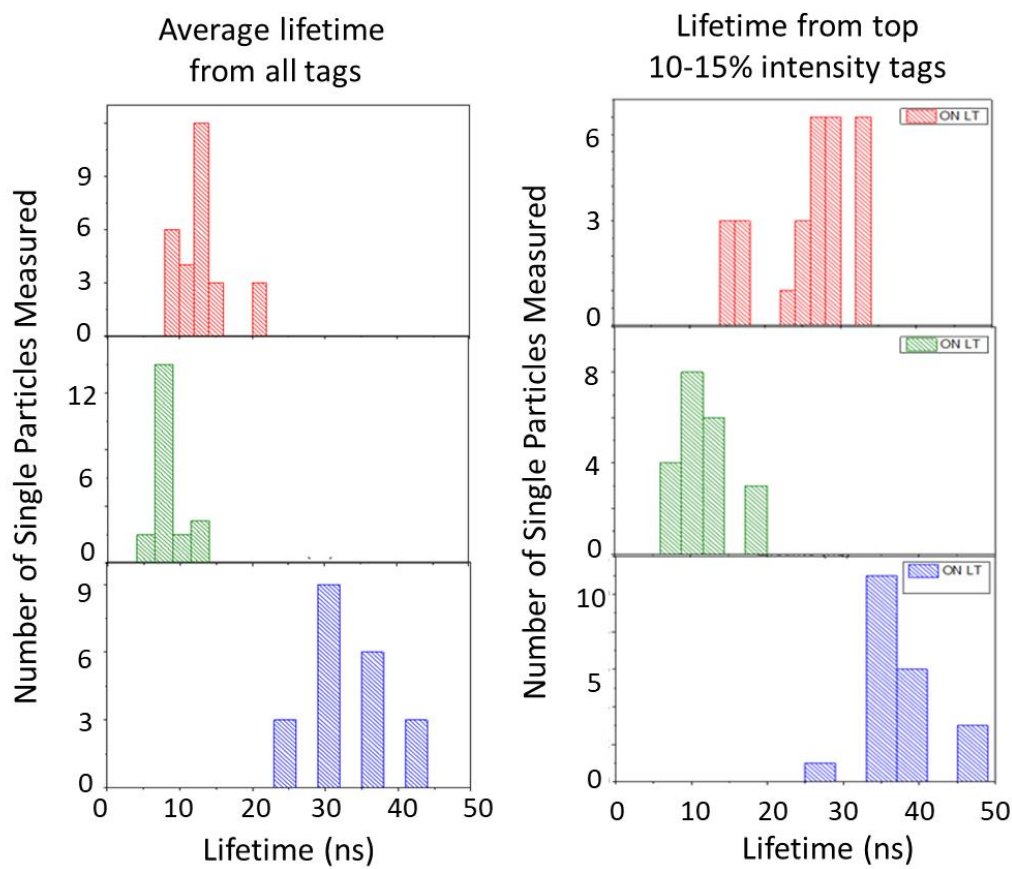

**Figure S10**. Statistical distribution of the single particle fluorescence lifetime (average) of various particles (monomer-blue; weak-neck CQDM- green; rod-like CQDM- red). Left column represents the histogram of average lifetimes consisting of all intensity tags during the measurements. In the right column, the lifetime only from the ON states (~top 15%) are plotted. The ON lifetime distribution exhibit a drastic shortening in weakly fused CQDMs, which shifts to longer values from neck-filled CQDMs. The distributions reflect the interplay between the neutral, and charged state emission as demonstrated in the FLID diagram in figure 3 and the discussions therein.

S11

**Section S8. Demonstration of Purity of the batch: Hydrodynamic size of the dimers:**

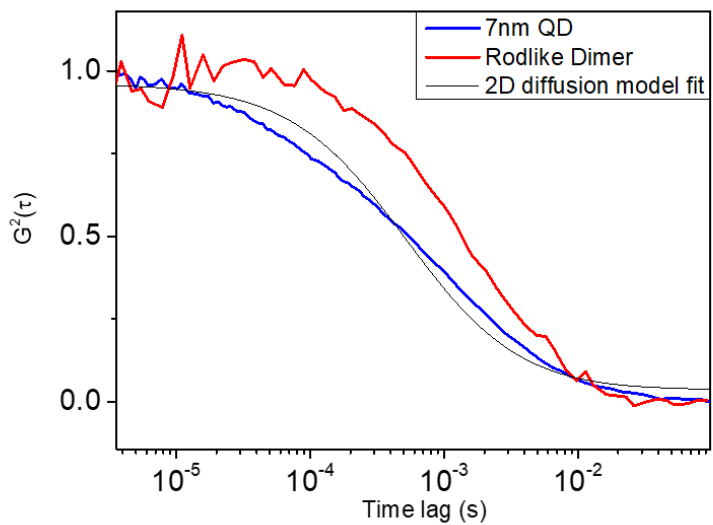

**Figure S11**. The purity of the dimer, tested with Fluorescence Correlation spectroscopy (FCS) of rod-like dimer sample in toluene. We measure the fluorescence fluctuation induced by the Brownian motion of the particle in solution[S4]. We measure the translational diffusion time of the particle in equilibrium with the solvent as a single particles diffuses in and out of the focal volume, leading to bright, and dark emission periods. The monomer (blue) possess deviation from the 2D diffusion model due to additional inherent blinking (while matching the hydrodynamic radius). The dimer shows nearly twice the diffusion time with reduced blinking signature in the FCS curve.

## Section S9. References: